\documentclass[a4paper]{article}
\usepackage{amsfonts}
\def\End{\mathop{\mathrm{End}}}
\def\Ad{\mathop{\mathrm{Ad}}}

\def\min{\mathop{\mathrm{min}}}
\def\id{\mathop{\mathrm{id}}}

\newtheorem{theorem}{Theorem}

\def\SURFACE{\Sigma}
\def\ISOTOP{\mathcal{S}}
\def\GENUS{g}
\def\VERTICES{V}
\def\NVERTICES{r}
\def\EULER{M}
\def\EDGES{E}
\def\TRIANGLES{T}
\def\REALS{\mathbb R}
\def\COMPLEXS{\mathbb C}
\def\PGROUP{\mathbb S}
\def\IT{i.t.}
\def\DIT{d.i.t.}
\def\SDIT{\Delta(\SURFACE)}
\def\HILBERT{L^2(\REALS)}
\def\HILBERTm{L^2(\REALS^m)}

\def\ROTATE{\mathsf R}
\def\PTOLEMY{\mathsf T}
\def\PERMUTE{\mathsf P}
\def\MOM{\mathsf p}
\def\POS{\mathsf q}
\def\QDILOG{\psi}
\def\IMUN{\mathsf i}
\def\FUNCTOR{\mathsf F}
\def\DEHN{\mathsf D}
\def\CONSTRAINT{\mathsf z}

\def\MCG{M_\SURFACE}
\begin{document}
\begin{titlepage}
\rightline{PDMI PREPRINT --- 24/1998}
\vspace{.5in}
\begin{center}
{\Large\bf Liouville central charge in quantum Teichm\"uller\\[6pt]
theory
}\\[0.5in]
\begin{tabular}{c}
{\large\bf R.\,M.\ Kashaev}\\[6pt]
Russian Academy of Sciences\\
Steklov Mathematical Institute\\
St. Petersburg Department\\
Fontanka 27, St. Petersburg 191011, Russia\\
E-mail:{\sf kashaev@pdmi.ras.ru}
\end{tabular}
\end{center}
\vspace{.5in}
\centerline{\large November 1998}
\vspace{.5in}
\centerline{\large ABSTRACT}
\vspace{.2in}
In the quantum Teichm\"uller theory, based
on Penner coordinates, the mapping class groups of punctured surfaces
are represented projectively. The case of a genus three surface with
one puncture is worked out explicitly. The projective factor
is calculated. It is given by the exponential of the Liouville central
charge.
\end{titlepage}

\section{Introduction}

In \cite{Faddeev} L.D. Faddeev
suggested to use
as a building block for the evolution operator in the quantum Liouville
model on discrete space-time \cite{Faddeev2}
the following non-compact version
of the quantum dilogarithm function:
    \begin{equation}\label{maluzh}
\QDILOG(x)\equiv\exp\left(-\frac{1}{4}\int_{-\infty}^{+\infty}
\frac{\exp(-\IMUN xz)\, dz}{\sinh(\pi\lambda z)
\sinh(\pi\lambda^{-1} z) z}\right),
    \end{equation}
where the singularity at $z=0$ is put below of the contour
of integration, and $\lambda$ is a real parameter.
Among other things this function satisfies an inversion relation
\begin{equation}\label{inv}
\QDILOG(x)\QDILOG(-x)=\exp\left(-\frac{\IMUN\pi}{12}
(\lambda^2+\lambda^{-2})-\frac{\IMUN x^2}{4\pi}\right).
\end{equation}
Faddeev conjectured that the inversion
factor in this equation (the constant part of the r.h.s)
is related to the Liouville central charge \cite{Faddeev1}.

The purpose of this paper is to prove Faddeev's conjecture.
Our approach is based on the quantum theory of Teichm\"uller
spaces of punctured surfaces suggested recently in \cite{Kashaev}
and independently by Fock, Chekhov and Frolov \cite{Fock1,Fock}.
It is known \cite{Verlinde} that the Teichm\"uller spaces appear
as the classical phase spaces
of Chern--Simons theory with gauge group $SL(2,\REALS)$.
This theory is related to
 $2+1$ dimensional gravity with cosmological constant
\cite{Achucarro,Witten,Verlinde}, which in turn induces the Liouville
theory at spatial infinity \cite{Carlip,Coussaert}.

The quantum theory of the Teichm\'uller spaces, described in \cite{Kashaev},
is based on Penner's coordinates \cite{Penner}.
The mapping class (or modular) group of a punctured surface is
projectively represented in the corresponding Hilbert space of
physical states in terms of the non-compact quantum dilogarithm.  We
work out explicitly the case of a genus three surface with one
puncture, and calculate the corresponding projective factor which,
being essentially given by the inversion factor of the non-compact
quantum dilogarithm, appears to be the exponential of the quantum
Liouville central charge.  This is in agreement with the
interpretation of the Hilbert space of quantum states as Virasoro
conformal blocks \cite{Verlinde}, since the projective factors in the
representations of the mapping class groups, coming from
two-dimensional conformal field theories, are known to be given by
the exponential of the corresponding Virasoro central charges
\cite{Friedan,Moore}.

\section{The non-compact quantum dilogarithm}
\subsection{Notation}

For any natural $m$ define embeddings
        \[
\iota_i\colon \End \HILBERT\ni a\mapsto a_i=1\otimes\cdots\otimes
a\otimes\cdots\otimes 1 \in\End\HILBERTm , \quad 1\le i\le m, \]
where $a$ stands in the $i$-th position, which mean that the operator
$a_i$ acts on the $i$-th argument as $a$, and identically on the
others:
\[
a_if(x_1,\ldots,x_m)=\left.
af(x_1,\ldots,x_{i-1},x,x_{i+1},\ldots,x_m)\right|_{x=x_i}.
\]
If $b\in \End L^2(\REALS^k)$ for some $0<k\le m$ and
$\{i_1,i_2,\ldots,i_k\}\subset\{1,2,\ldots,m\}$,
we write
    \[
b_{i_1i_2\ldots i_2}\equiv\iota_{i_1}\otimes\iota_{i_2}\otimes\cdots
\otimes\iota_{i_k}(b).
    \]
The permutation group $\PGROUP_m$
is represented faithfully in $\End\HILBERTm$:
    \begin{equation}\label{perm}
\PERMUTE_\sigma f(\ldots,x_i,\ldots)\equiv f(\ldots,x_{\sigma(i)},\ldots),
\quad \forall\sigma\in \PGROUP_m.
    \end{equation}

\subsection{Some algebraic relations}

Let $\MOM,\POS\in \End\HILBERT$ be defined by
    \[
\MOM f(x)\equiv -
2\pi\IMUN\partial f(x)/\partial x,\quad
\POS f(x)\equiv xf(x).
    \]
They satisfy the Heisenberg commutation relation
    \[
[\POS,\MOM]=2\pi\IMUN.
    \]
The non-compact quantum dilogarithm, defined by eqn~(\ref{maluzh}),
solves a pair of functional equations
    \[
\QDILOG(x+\IMUN\pi\lambda^{\pm1})=(1+\exp(x\lambda^{\pm1}))
\QDILOG(x-\IMUN\pi\lambda^{\pm1}),
    \]
and, as was argued in \cite{Faddeev}, satisfies the following
five-term operator equation:
\begin{equation}\label{5-term}
\QDILOG(\POS)\QDILOG(\MOM)=\QDILOG(\MOM)\QDILOG(\MOM+\POS)\QDILOG(\POS).
    \end{equation}
This fact can be proved rigorously \cite{Woronowicz}.

We consider a unitary operator
$\PTOLEMY\in\End L^2(\REALS^2)$, defined by
    \begin{equation}
\PTOLEMY\equiv\exp\left(\frac{\IMUN}{2\pi}\MOM_1\POS_2\right)
\QDILOG(\POS_1+\MOM_2-\POS_2).
    \end{equation}
Eqn~(\ref{5-term}) is equivalent to
    \begin{equation}\label{pentagon}
\PTOLEMY_{12}\PTOLEMY_{13}\PTOLEMY_{23}=\PTOLEMY_{23}\PTOLEMY_{12},
    \end{equation}
while the inversion relation~(\ref{inv}) leads to the equation
    \begin{equation}\label{inversion}
\PTOLEMY_{12}\ROTATE_1\PTOLEMY_{21}\ROTATE_1^{-1}=
\zeta\ROTATE_1\PERMUTE_{(12)},
    \end{equation}
where unitary operator $\ROTATE\in\End\HILBERT$ is defined by
    \begin{equation}\label{rotation}
\ROTATE f(x)
\equiv\exp\left(\frac{\IMUN x^2}{4\pi}-\frac{\IMUN\pi}{12}
\right)\int_{-\infty}^{+\infty}
 f(y)\exp\left(\frac{\IMUN xy}{2\pi}\right)\frac{dy}{2\pi},
    \end{equation}
and
    \begin{equation}\label{prfac}
\zeta=\exp\left(-\frac{\IMUN\pi}{12}(\lambda+\lambda^{-1})^2\right).
    \end{equation}
 Besides these we have a symmetry property
    \begin{equation}\label{symmetry}
\PTOLEMY_{12}=\ROTATE_1^{-1}\ROTATE_2\PTOLEMY_{21}\ROTATE_1
\ROTATE_2^{-1}
    \end{equation}
with the same operator $\ROTATE$ defined by eqn~(\ref{rotation}).
Note that $\ROTATE$ satisfies the following equations:
\[
\ROTATE\POS\ROTATE^{-1}=\MOM-\POS,\quad
\ROTATE\MOM\ROTATE^{-1}=-\POS,\quad
\ROTATE^3=1.
    \]
It can also be written in an operator form
    \[
\ROTATE=\exp\left(-\frac{\IMUN\pi}{3}\right)
\exp\left(\frac{\IMUN}{2\pi}\POS^2\right)
\exp\left(\frac{\IMUN}{4\pi}\MOM^2\right)
\exp\left(\frac{\IMUN}{4\pi}\POS^2\right).
    \]

\section{Projective representations of mapping class groups}

We call a two-cell in CW complex {\em triangle} if exactly
three boundary points of the corresponding two-disk
are mapped to the zero-skeleton. We shall also call
zero-cells and one-cells {\em vertices} and {\em edges}, respectively.

Let $\SURFACE=\SURFACE_{\GENUS,\NVERTICES}$ be a closed oriented
Riemann surface
of genus $\GENUS$ with a set $\VERTICES$ of $\NVERTICES>0$
marked points such that $\EULER\equiv 2\GENUS-2+\NVERTICES>0$.
By {\em ideal triangulation (\IT)} $\tau$ of $\SURFACE$ we mean
the isotopy class of a representation
of $\SURFACE$ as a CW complex with $\VERTICES$
as the set of vertices,
and all two-cells being triangles. It is easy to count that
there are $2\EULER$ triangles and $3\EULER$ edges in any \IT.
We shall denote by $\EDGES(\tau)$ and $\TRIANGLES(\tau)$ the sets of
edges and triangles respectively in \IT~$\tau$.

An \IT~$\tau$ is called {\em decorated (\DIT)} if each triangle is provided
by a marked corner and a bijective {\em ordering} mapping
    \[
\bar\tau\colon
\{1,\ldots,2\EULER\}\ni j\mapsto\bar\tau_j\in\TRIANGLES(\tau)
    \]
is fixed. Denote the set of all \DIT{} as $\SDIT$.
The permutation group $\PGROUP_{2\EULER}$ naturally
acts in $\SDIT$ from the right:
    \[
\SDIT\times\PGROUP_{2\EULER}\ni(\tau,\sigma)\mapsto\tau\sigma
\in\SDIT,
    \]
where $\tau\sigma$ differs from $\tau$ only in ordering:
    \[
\overline{\tau\sigma}=\bar\tau\circ\sigma.
    \]
We also define two elementary
transformations
\[
\rho^t,\omega_e\colon\SDIT\to\SDIT
\]
as follows.

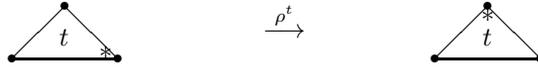
\begin{figure}[ht]
\centering
\begin{picture}(200,40)
\put(0,0){\begin{picture}(40,20)
\put(0,0){\line(1,0){40}}
\put(0,0){\line(1,1){20}}
\put(20,20){\line(1,-1){20}}
\put(0,0){\circle*{3}}
\put(20,20){\circle*{3}}
\put(40,0){\circle*{3}}
\put(33,0){$*$}
\put(18,5){$t$}
\end{picture}}
\put(160,0){\begin{picture}(40,20)
\put(0,0){\line(1,0){40}}
\put(0,0){\line(1,1){20}}
\put(20,20){\line(1,-1){20}}
\put(0,0){\circle*{3}}
\put(20,20){\circle*{3}}
\put(40,0){\circle*{3}}
\put(17.5,14){$*$}
\put(18,5){$t$}
\end{picture}}
\put(95,8){$\stackrel{\rho^t}{\longrightarrow}$}
\end{picture}
\caption{Elementary transformation $\rho^t$ consisting of changing the
marked corner of triangle $t$.}\label{ft}
\end{figure}

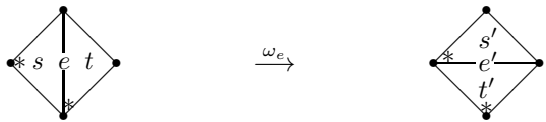
\begin{figure}[ht]
\centering
\begin{picture}(200,40)
\put(0,0){
\begin{picture}(40,40)
\put(20,0){\line(-1,1){20}}
\put(40,20){\line(-1,-1){20}}
\put(0,20){\line(1,1){20}}
\put(40,20){\line(-1,1){20}}
\put(20,0){\line(0,1){17}}
\put(20,40){\line(0,-1){16}}
\put(18,18){$e$}
\put(8,18){$s$}\put(28,18){$t$}
\put(20,0){\circle*{3}}
\put(0,20){\circle*{3}}
\put(20,40){\circle*{3}}
\put(40,20){\circle*{3}}
\put(1,18){$*$}
\put(19.5,2){$*$}
\end{picture}}
\put(160,0){
\begin{picture}(40,40)
\put(20,0){\line(-1,1){20}}
\put(40,20){\line(-1,-1){20}}
\put(0,20){\line(1,1){20}}
\put(40,20){\line(-1,1){20}}
\put(40,20){\line(-1,0){17}}
\put(0,20){\line(1,0){16}}
\put(17,17){$e'$}
\put(17,26){$s'$}\put(17,7){$t'$}
\put(20,0){\circle*{3}}
\put(0,20){\circle*{3}}
\put(20,40){\circle*{3}}
\put(40,20){\circle*{3}}
\put(3,20){$*$}
\put(17.5,1){$*$}
\end{picture}}
\put(95,17){$\stackrel{\omega_e}{\longrightarrow}$}
\end{picture}
\caption{Elementary transformation $\omega_e$ consisting of replacing
edge $e$ by the complementary edge $e'$.}\label{fe}
\end{figure}

Let $t$ be a triangle in \DIT~$\tau$.
We define \DIT~$\tau^t\equiv\rho^t(\tau)$
by changing the marked corner of triangle $t$ as is shown
in Figure~\ref{ft} and leaving the ordering mapping unchanged:
$\bar\tau=\bar\tau^t$.
Let now $e$ be an edge in \DIT~$\tau$
shared by two distinct triangles $s$, $t$ with marked
corners as is shown in the l.h.s. of Figure~\ref{fe}.
 Then \DIT~$\tau_e\equiv \omega_e(\tau)$
is obtained from $\tau$ by replacing the cells $e,\ s,\ t$
by new cells $e',\ s',\ t'$ as is shown in the r.h.s. of
Figure~\ref{fe}. The ordering mapping $\bar\tau_e$ is defined by
    \[
\bar\tau_e^{-1}\left|_{\TRIANGLES(\tau_e)-\{s',t'\}}=
\bar\tau^{-1}\right|_{\TRIANGLES(\tau)-\{s,t\}},\quad
\bar\tau_e^{-1}(s')=\bar\tau^{-1}(s).
    \]

For each \DIT~$\tau$ and for each $1\le i\le 2\EULER$ define
    \[
\FUNCTOR(\tau,\tau^{\bar\tau(i)})\equiv\ROTATE_i\in
\End L^2(\REALS^{2\EULER}).
    \]
Let $i\ne j$ be such that triangles
$s\equiv\bar\tau(i)$ and $t\equiv\bar\tau(j)$ share an edge $e$ as in
the l.h.s. of Figure~\ref{fe}. Then we define
    \begin{equation}\label{tij}
\FUNCTOR(\tau,\tau_e)\equiv\PTOLEMY_{ij}\in
\End L^2(\REALS^{2\EULER}).
    \end{equation}
Finally, for any permutation
$\sigma\in\PGROUP_{2\EULER}$ define
    \[
\FUNCTOR(\tau,\tau\sigma)\equiv\PERMUTE_\sigma,
    \]
where $\PERMUTE_\sigma$ is defined by eqn~(\ref{perm}).
We claim that due to relations~(\ref{pentagon}),
(\ref{symmetry}) and (\ref{inversion}),
$\FUNCTOR$ can be extended to a unitary operator
valued function
$\FUNCTOR(\tau,\tau')$ on $\SDIT\times\SDIT$
such that for any \DIT~$\tau$, $\tau'$, $\tau''$
\[
 \FUNCTOR(\tau,\tau')\FUNCTOR(\tau',\tau'')\FUNCTOR(\tau'',\tau)\in
\COMPLEXS.
\]

The mapping class or modular group $\MCG$ of $\SURFACE$ naturally acts
in $\SDIT$.
By construction we have the following invariance property
of the function $\FUNCTOR$:
\[
 \FUNCTOR(f(\tau),f(\tau'))=
 \FUNCTOR(\tau,\tau'),\qquad \forall f\in\MCG.
\]
This enables us to construct a unitary projective
representation of $\MCG$:
    \[
\MCG\ni f\mapsto\FUNCTOR(\tau,f(\tau))\in
\End L^2(\REALS^{2\EULER}).
    \]
Indeed,
\[
\FUNCTOR(\tau,f(\tau))\FUNCTOR(\tau,h(\tau))=
\FUNCTOR(\tau,f(\tau))\FUNCTOR(f(\tau),f(h(\tau)))
\simeq\FUNCTOR(\tau,fh(\tau)),
\]
where we denote by $\simeq$ an equality up to a numerical factor.
In the next section we calculate the projective factor in the case
of the surface $\SURFACE_{3,1}$ of genus three and with one puncture.

\section{The case of $\SURFACE_{3,1}$}

To find the central charge we shall need a suitable presentation
of the mapping class group. First, following \cite{Luo}
introduce some notation.

\begin{figure}[ht]
\centering
\begin{picture}(120,40)
\put(0,20){\line(1,0){40}}
\put(20,0){\line(0,1){40}}
\put(100,20){\line(1,0){10}}
\put(130,20){\line(1,0){10}}
\put(120,0){\line(0,1){10}}
\put(120,30){\line(0,1){10}}
\put(110,20){\line(1,1){10}}
\put(120,10){\line(1,1){10}}
\put(60,18){$\longrightarrow$}
\put(-6,18){$a$}
\put(21,34){$b$}
\put(21,15){$p$}
\put(125,10){$ab$}
\end{picture}
\caption{Resolution from $a$ to $b$ at $p$.}\label{fresolution}
\end{figure}
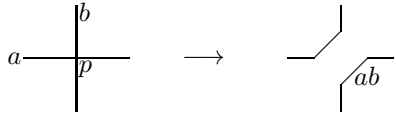

Let $\ISOTOP=\ISOTOP(\SURFACE)$ be the set of isotopy classes of
simple closed curves on the surface $\SURFACE$. We denote by $D_\alpha$,
$\alpha\in\ISOTOP$, the Dehn twist along $\alpha$. Define
\[
I(\alpha,\beta)=\min\{|a\cap b|\colon a\in\alpha,b\in\beta\},\quad
\forall\alpha,\beta\in\ISOTOP.
\]
We shall use $\alpha\cap\beta=\emptyset$ to denote $I(\alpha,\beta)=0$;
$\alpha\perp\beta$ to denote $I(\alpha,\beta)=1$; and
$\alpha\perp_0\beta$ to denote $I(\alpha,\beta)=2$ so that their
algebraic intersection number is zero. If $a,b$ are two arcs intersecting
transversely at a point $p$, then the {\em resolution of $a\cup b$ at $p$
from $a$ to $b$} is defined as follows. Fix any orientation on $a$
and use the orientation on the surface to determine an orientation
on $b$. Then resolve the intersection according to the orientations,
see Figure~\ref{fresolution}.

The resolution is independent on the choice
of the orientations on $a$. If $\alpha\perp\beta$ or $\alpha\perp_0\beta$,
take $a\in\alpha,b\in\beta$ so that $|a\cap b|=I(\alpha,\beta)$.
Then the curve obtained by resolving all intersection points in $a\cap b$
from $a$ to $b$ is again simple closed curve. Denote by
$\alpha\beta$ its' isotopy class. It follows that when $\alpha\perp\beta$
then $\alpha\beta\perp\alpha,\beta$ and
when $\alpha\perp_0\beta$
then $\alpha\beta\perp_0\alpha,\beta$. Also the Dehn twist of $\beta$
along $\alpha$ is given by $D_\alpha(\beta)=\alpha\beta$ if
$\alpha\perp\beta$, and $D_\alpha(\beta)=\alpha(\alpha\beta)$ if
$\alpha\perp_0\beta$. Let $N(a)$ and $N(b)$ be small regular neighborhoods
of $a$ and $b$. Then $N(a\cup b)=N(a)\cup N(b)$ is homeomorphic
to a torus with one boundary component when $\alpha\perp\beta$ , and
to a sphere with four boundary components when $\alpha\perp_0\beta$.
Denote by $\partial(\alpha,\beta)$ the isotopy class of the curve system
$\partial N(a\cup b)$.

One has a {\em lantern} relation \cite{Dehn,Johnson}:
\begin{equation}\label{lantern}
D_\alpha D_\beta D_{\alpha\beta}=
D_{\partial(\alpha,\beta)}\mbox{ if }\alpha\perp_0\beta,
\end{equation}
and a {\em chain} relation \cite{Dehn}:
\begin{equation}\label{chain}
(D_\alpha D_\beta D_\gamma)^4=D_{\epsilon_1}D_{\epsilon_2}
\mbox{ if }\alpha,\beta,\gamma,\epsilon_i
\mbox{ are as shown in Figure \ref{fchain}.}
\end{equation}
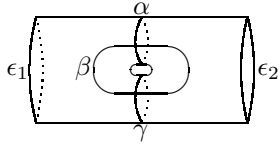
\begin{figure}[ht]
\centering
\begin{picture}(90,40)
\put(5,40){\line(1,0){80}}
\put(5,0){\line(1,0){80}}
\put(45,20){\oval(8,4)}
\put(45,20){\oval(36,18)}
\qbezier(45,22)(40,30)(45,40)
\qbezier[7](45,22)(50,30)(45,40)
\qbezier(45,18)(40,10)(45,0)
\qbezier[7](45,18)(50,10)(45,0)
\qbezier(85,0)(80,20)(85,40)
\qbezier(85,0)(90,20)(85,40)
\qbezier(5,0)(0,20)(5,40)
\qbezier[15](5,0)(10,20)(5,40)
\put(-6,18){$\epsilon_1$}
\put(20,18){$\beta$}
\put(89,18){$\epsilon_2$}
\put(42,-5){$\gamma$}
\put(42,41){$\alpha$}
\end{picture}
\caption{The Dehn twists along these curves satisfy the chain relation.}
\label{fchain}
\end{figure}
\begin{theorem}[Gervais \cite{Gervais}]
For any compact oriented surface $\SURFACE$ of genus $\GENUS>1$, the mapping
class group $\MCG$ has the following presentation:
\[
\begin{array}{rcl}
 \mbox{generators}\colon &&\{
D_\alpha\colon\mbox{ non-separating }\alpha
\in\ISOTOP(\SURFACE)\};\\
 \mbox{relations}\colon&(i)&
D_\alpha D_\beta=D_\beta D_\alpha\mbox{ if }\alpha\cap\beta
=\emptyset,\\
&(ii)&
D_{\alpha\beta}=D_\alpha D_\beta D_\alpha^{-1}\mbox{ if }
\alpha\perp\beta,\\
&(iii)&\mbox{one lantern relation (\ref{lantern})},\\
&(iv)&\mbox{one chain relation (\ref{chain})}.
\end{array}
\]
\end{theorem}
If $\alpha \perp\beta$, relations $(iii)$ together with
$\alpha\beta\perp\alpha$, $(\alpha\beta)\alpha=\beta$ imply that
\begin{equation}\label{braid}
D_\alpha D_\beta D_\alpha=D_\beta D_\alpha D_\beta \mbox{ if } \alpha\perp
\beta.
\end{equation}

\subsection{Realization of the Dehn twists}

\begin{figure}[ht] \centering
\begin{picture}(320,40)
\put(0,0){
\begin{picture}(80,40)
\put(40,20){\oval(8,4)}
\put(40,20){\oval(80,40)}
\put(0,20){\circle*{3}}
\put(44,20){\circle*{3}}
\qbezier(0,20)(50,5)(44,20)
\qbezier(0,20)(50,35)(44,20)
\put(0.5,14){$*$}
\put(39,22){$*$}
\put(18,17.5){$\bar\tau_0$}
\put(65,17.5){$\bar\tau_1$}
{\thicklines
\put(40,20){\circle{25}}
}
\put(53,17.5){$\alpha$}
\end{picture}
}
\put(95,18){$\stackrel{D_\alpha}{\longrightarrow}$}
\put(120,0){
\begin{picture}(80,40)
\put(40,20){\oval(8,4)}
\put(40,20){\oval(80,40)}
\put(0,20){\circle*{3}}
\put(44,20){\circle*{3}}
\put(38.5,20){\oval(11,10)[b]}
\put(38.5,20){\oval(11,10)[lt]}
\qbezier(38.5,25)(65,26)(65,20)
\qbezier(0,20)(68,-15)(65,20)
\qbezier(0,20)(50,5)(44,20)
\put(45,17.5){$*$}
\put(0.5,13.5){$*$}
\put(20,27){$\bar\tau'_1$}
\put(49,10){$\bar\tau'_0$}
\put(60.5,24){$e$}
\end{picture}
}
\put(215,18){$\stackrel{\omega_e}{\longrightarrow}$}
\put(240,0){
\begin{picture}(80,40)
\put(40,20){\oval(8,4)}
\put(40,20){\oval(80,40)}
\put(0,20){\circle*{3}}
\put(44,20){\circle*{3}}
\qbezier(0,20)(50,5)(44,20)
\qbezier(0,20)(50,35)(44,20)
\put(0.5,14){$*$}
\put(39,22){$*$}
\put(18,17.5){$\bar\tau_0$}
\put(65,17.5){$\bar\tau_1$}
\end{picture}
}
\end{picture}
\caption{The Dehn twist $\tau'\equiv D_\alpha(\tau)$
 of d.i.t.~$\tau$
of an annulus along contour $\alpha$ is mapped by the elementary
transformation $\omega_e$ back to $\tau=\omega_e(\tau')$.}\label{fannulus}
\end{figure}
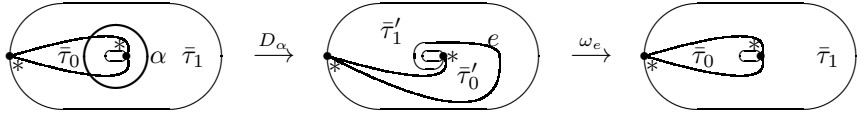

To begin with\footnote{In the rest of the paper we numerate
triangles from $0$ thru $2\EULER-1$ rather than from $1$ thru
$2\EULER$.}, we describe the simplest case of the Dehn twist of an
annulus along the only simple non-contractible curve denoted in
Figure~\ref{fannulus} as $\alpha$.  
From the same Figure~\ref{fannulus} it follows that
$\omega_e\circ D_\alpha=\id$.
So, using definition (\ref{tij}), we obtain
\begin{equation}\label{t01}
\FUNCTOR(\tau,D_\alpha(\tau))=\FUNCTOR(D_\alpha^{-1}(\tau),\tau)=
\FUNCTOR(\omega_e(\tau),\tau)\simeq\PTOLEMY_{01}^{-1},
\end{equation}
where the normalization remains to be fixed.

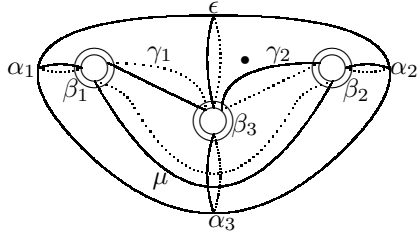
\begin{figure}[ht]
\centering
\begin{picture}(200,120)
\put(112,83){\circle*{3}}
\qbezier(50,50)(100,0)(150,50)
\qbezier(50,50)(0,100)(100,100)
\qbezier(150,50)(200,100)(100,100)
\put(55,80){\circle{10}}
\put(145,80){\circle{10}}
\put(100,60){\circle{10}}
\qbezier(100,100)(95,85)(100,65)
\qbezier[20](100,100)(105,85)(100,65)
\put(98,101){$\epsilon$}
\put(55,80){\circle{15}}
\put(43,69){$\beta_1$}
\put(145,80){\circle{15}}
\put(149,69){$\beta_2$}
\put(100,60){\circle{15}}
\put(107,56){$\beta_3$}
\qbezier(103,64)(103,82)(140,82)
\qbezier[20](103,64)(120,72)(140,82)
\put(120,84){$\gamma_2$}
\qbezier(97,64)(80,72)(60,82)
\qbezier[20](97,64)(97,82)(60,82)
\put(75,84){$\gamma_1$}
\qbezier(55,75)(100,-5)(145,75)
\qbezier[40](70,65)(100,15)(130,65)
\qbezier[10](70,65)(67,72)(55,75)
\qbezier[10](145,75)(133,72)(130,65)
\put(77,36){$\mu$}
\qbezier(100,55)(95,40)(100,25)
\qbezier[25](100,55)(105,40)(100,25)
\put(98,20){$\alpha_3$}
\qbezier(50,80)(42,83)(34,80)
\qbezier[10](50,80)(42,77)(34,80)
\put(22,78){$\alpha_1$}
\qbezier(150,80)(158,83)(166,80)
\qbezier[10](150,80)(158,77)(166,80)
\put(167,78){$\alpha_2$}
\end{picture}
\caption{The Dehn twists along these curves generate the mapping
class group of the surface $\Sigma_{3,1}$}\label{fcurves}
\end{figure}

Now we consider the curves on $\SURFACE_{3,1}$, shown in
Figure~\ref{fcurves}.
We choose the \DIT~$\tau$, obtained by cutting the surface into
three handles and a triangle as shown in Figures~\ref{fcut},
\ref{fhandles}.
\begin{figure}[ht]
\centering
\begin{picture}(200,120)
\put(112,83){\circle*{3}}
\qbezier(50,50)(100,0)(150,50)
\qbezier(50,50)(0,100)(100,100)
\qbezier(150,50)(200,100)(100,100)
\put(55,80){\circle{10}}
\put(145,80){\circle{10}}
\put(100,60){\circle{10}}
\qbezier(50,50)(86,83)(112,83)
\qbezier(80,99)(100,89)(112,83)
\qbezier[40](80,99)(65,75)(50,50)
\put(72,80){$c_1$}
\qbezier(112,83)(116,91)(120,99)
\qbezier(112,83)(130,70)(150,50)
\qbezier[40](120,99)(135,75)(150,50)
\put(129,87){$c_2$}
\qbezier(112,83)(83,80)(80,29)
\qbezier(120,29)(118,80)(112,83)
\qbezier[50](120,29)(100,108)(80,29)
\put(85,40){$c_3$}
\end{picture}
\caption{The surface $\Sigma_{3,1}$ is cut into a triangle and
three handles}\label{fcut}
\end{figure}
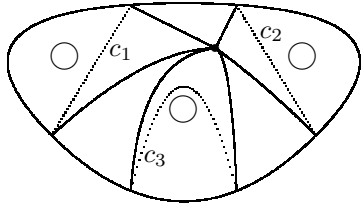
\begin{figure}[ht]
\centering
\begin{picture}(200,40)
\put(0,0){
\begin{picture}(100,40)
\put(0,0){\line(1,0){100}}
\put(100,0){\line(0,1){40}}
\put(100,40){\line(-1,0){100}}
\put(0,0){\line(0,1){40}}
\put(0,0){\circle*{3}}
\put(100,0){\circle*{3}}
\put(100,40){\circle*{3}}
\put(0,40){\circle*{3}}
\qbezier(100,40)(90,36)(80,36)
\qbezier(100,40)(90,36)(90,32)
\put(80,32){\oval(20,8)[l]}
\put(80,32){\oval(20,8)[br]}
\put(71,30.5){$c_i$}
\qbezier(0,0)(15,36)(100,40)
\qbezier(0,0)(85,8)(100,40)
\put(3,1){$*$}
\put(94,1){$*$}
\put(1,34){$*$}
\put(8,29){$\bar\tau_{3i-1}$}
\put(37,18){$\bar\tau_{3i-2}$}
\put(79,9){$\bar\tau_{3i}$}
\end{picture}}
\put(152,0){\begin{picture}(48,40)
\put(0,0){\line(3,5){24}}
\put(0,0){\line(1,0){48}}
\put(48,0){\line(-3,5){24}}
\put(0,0){\circle*{3}}
\put(48,0){\circle*{3}}
\put(24,40){\circle*{3}}
\put(20,14){$\bar\tau_0$}
\put(21.5,32){$*$}
\put(20,2.5){$c_3$}
\put(2.5,20){$c_2$}
\put(36,20){$c_1$}
\end{picture}}
\end{picture}
\caption{
D.i.t.~$\tau$ of the surface $\Sigma_{3,1}$.
Index $i$ takes three values $1,2,3$.
The rectangles with identified opposite sides represent
the three handles, which are glued along the boundary loops $c_i$
to the corresponding sides of the triangle $\bar\tau_0$.}
\label{fhandles}
\end{figure}
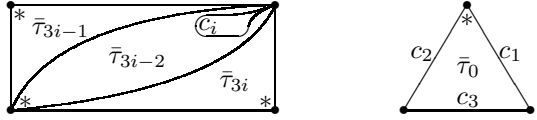
The corresponding Dehn twists can be calculated by choosing appropriate
\DIT{} where the annular neighborhoods of the curves look
like as in Figure~\ref{fannulus},
and using formula~(\ref{t01}). The result of such calculations reads:
\[
\begin{array}{l}
\FUNCTOR_{\alpha_i}=\zeta^{-6}\Ad(\PTOLEMY_{3i-1,3i-2})
\PTOLEMY_{3i,\widehat{3i-2}}^{-1},\\
\FUNCTOR_{\beta_i}=\zeta^{-6}\Ad(\PTOLEMY_{3i,\widehat{3i-2}})
\PTOLEMY_{\widehat{3i-1},3i}^{-1},
\end{array}
\quad i=1,2,3;
\]
\[
\FUNCTOR_{\gamma_1}=\zeta^{-6}\Ad(\PTOLEMY_{0\check7}\PTOLEMY_{87}
\PTOLEMY_{10}\PTOLEMY_{1\check9}\PTOLEMY_{21}
\PTOLEMY_{\check31}\PTOLEMY_{27}\PTOLEMY_{30})
\PTOLEMY_{\check0\hat7}^{-1},
\]
\[
\FUNCTOR_{\gamma_2}=\zeta^{-6}\Ad(\PTOLEMY_{87}\PTOLEMY_{8\check0}
\PTOLEMY_{8\check4}\PTOLEMY_{\hat89}\PTOLEMY_{69}
\PTOLEMY_{\check97}\PTOLEMY_{54}\PTOLEMY_{\check78})
\PTOLEMY_{\check8\hat4}^{-1},
\]
\[
\FUNCTOR_{\mu}=\zeta^{-6}\Ad(\PTOLEMY_{0\check7}\PTOLEMY_{87}
\PTOLEMY_{10}\PTOLEMY_{1\check9}\PTOLEMY_{\hat89}
\PTOLEMY_{21}\PTOLEMY_{49}\PTOLEMY_{69}\PTOLEMY_{54}\PTOLEMY_{5\check8}
\PTOLEMY_{08}\PTOLEMY_{30}\PTOLEMY_{3\check4}\PTOLEMY_{\hat49}
\PTOLEMY_{\hat10})\PTOLEMY_{\check1\hat9}^{-1},
\]
\[
\FUNCTOR_{\alpha_3\mu}=\zeta^{-6}\Ad(\PTOLEMY_{0\check7}\PTOLEMY_{87}
\PTOLEMY_{10}\PTOLEMY_{\hat89}\PTOLEMY_{18}
\PTOLEMY_{21}\PTOLEMY_{49}\PTOLEMY_{69}\PTOLEMY_{30}\PTOLEMY_{\check97}
\PTOLEMY_{4\check2}\PTOLEMY_{54}\PTOLEMY_{\hat47}\PTOLEMY_{\hat10}
\PTOLEMY_{50})\PTOLEMY_{\check0\hat7}^{-1},
\]
\[
\FUNCTOR_{\epsilon}=\zeta^{-6}\Ad(\PTOLEMY_{0\check7}\PTOLEMY_{87}
\PTOLEMY_{10}\PTOLEMY_{30}\PTOLEMY_{\check20}
\PTOLEMY_{\hat30}\PTOLEMY_{21}\PTOLEMY_{\hat32}\PTOLEMY_{93}
\PTOLEMY_{\check31}
\PTOLEMY_{\check10})\PTOLEMY_{\check0\hat7}^{-1},
\]
where
\[
\FUNCTOR_{\alpha}\simeq\FUNCTOR(\tau,D_{\alpha}(\tau)),
\]
\[
\Ad(a)b\equiv aba^{-1},
\]
\[
a_{\hat k}\equiv \Ad(\ROTATE_k)a_k,\quad
a_{\check k}\equiv \Ad(\ROTATE_k^{-1})a_k,
\]
see eqn~(\ref{rotation}) for the definition of the operator
$\ROTATE$.
Note that the mutual normalization of these operators is fixed by the
relations of the type (\ref{braid}):
\[
\FUNCTOR_{\alpha_1}\FUNCTOR_{\beta_1}\FUNCTOR_{\alpha_1}=
\FUNCTOR_{\beta_1}\FUNCTOR_{\alpha_1}\FUNCTOR_{\beta_1},\quad\ldots
\]
while the overall normalization is fixed by the lantern relation:
\[
\FUNCTOR_{\alpha_3}\FUNCTOR_{\mu}\FUNCTOR_{\alpha_3\mu}=
\FUNCTOR_{\alpha_1}\FUNCTOR_{\alpha_2}\FUNCTOR_{\gamma_1}
\FUNCTOR_{\gamma_2}.
\]
Thus, we do not have any other freedom in normalization. Checking
the chain relation we recover the projective factor:
\begin{equation}\label{qchain}
(\FUNCTOR_{\alpha_1}\FUNCTOR_{\beta_1}\FUNCTOR_{\gamma_1})^4=
\xi_{\FUNCTOR}\FUNCTOR_{\epsilon}\FUNCTOR_{\alpha_3},\quad
\xi_{\FUNCTOR}=\zeta^{-72}.
\end{equation}

\subsection{Relation to the Liouville central charge}

In the projective representations of the mapping class groups,
associated with quantum CFT, the projective factors have the form
$\exp(2\pi\IMUN nc /24)$, where $n$ is an integer and
$c$ is the Virasoro central charge \cite{Friedan,Moore}.
According to the result of \cite{Kashaev} on quantization
of the Teichm\"uller spaces of punctured surfaces, the representation
given by the operators $\FUNCTOR_\alpha$ is not quite the representation,
corresponding to the quantum Teichm\"uller space of $\Sigma_{3,1}$.
There are additional degrees of
freedom, given by the constraints associated with homologies of the surface.
Explicit form of six basis constraints, corresponding to oriented
contours $\alpha_i,\beta_i$, is as follows:
\[
\begin{array}{l}
\CONSTRAINT_{\alpha_i}=\POS_{3i-2}-\MOM_{3i-2}+\MOM_{3i-1}-\MOM_{3i},\\
\CONSTRAINT_{\beta_i}=\MOM_{3i-2}-\POS_{3i-2}-\POS_{3i-1}+\POS_{3i},
\end{array}
\quad
i=1,2,3.
\]
All the constraints, corresponding to other contours, are linear
combinations of the basis ones:
\[
\CONSTRAINT_{\gamma_i}=\CONSTRAINT_{\alpha_i}-\CONSTRAINT_{\alpha_3},\quad
i=1,2,
\]
\[
\CONSTRAINT_{\mu}=\CONSTRAINT_{\alpha_1}-\CONSTRAINT_{\alpha_2},\quad
\CONSTRAINT_{\alpha_3\mu}=\CONSTRAINT_{\alpha_3}-
\CONSTRAINT_{\alpha_1}-\CONSTRAINT_{\alpha_2},\quad
\CONSTRAINT_{\epsilon}=\CONSTRAINT_{\alpha_3}.
\]
To get rid of these degrees of freedom, we note the following action
of the Dehn twists on the constraints:
\begin{equation}\label{contrans}
\Ad(\FUNCTOR_\alpha)\CONSTRAINT_\beta=\CONSTRAINT_\beta+
\alpha\circ\beta\CONSTRAINT_\alpha,
\end{equation}
for oriented contours $\alpha,\beta$, $\alpha\circ\beta$ being their
algebraic intersection index. The constraints themselves satisfy
the following commutation relations:
\begin{equation}\label{conalg}
[\CONSTRAINT_\alpha,\CONSTRAINT_\beta]=4\pi\IMUN\alpha\circ\beta.
\end{equation}
From eqns~(\ref{contrans}), (\ref{conalg}) it follows that the combinations
\[
\DEHN_\alpha\equiv\exp(\IMUN\CONSTRAINT_\alpha^2/8\pi)\FUNCTOR_\alpha
\]
commute with the constraints and satisfy the defining relations for the
Dehn twists with the projective factor $\xi_{\DEHN}=-\xi_{\FUNCTOR}$
in eqn~(\ref{qchain}). Taking into account eqn~(\ref{prfac}) we
obtain
\begin{equation}\label{eqproj}
\xi_{\DEHN}=-\xi_{\FUNCTOR}=-\zeta^{-72}=\exp(\IMUN\pi c_L),
\end{equation}
where
\begin{equation}\label{eqcharge}
c_L=1+6(\lambda+\lambda^{-1})^2\pmod 2
\end{equation}
is the Virasoro central charge in quantum Liouville theory. This is
in agreement with interpretation of physical states of quantum
Teichm\"uller theory as Virasoro conformal blocks.

\section{Summary}
In the quantum Teichm\"uller theory
of punctured surfaces, based on Penner coordinates,
the mapping class groups are represented projectively in terms
of non-compact quantum dilogarithm. Algebraically the representations
are based on three operator equations~(\ref{pentagon}),(\ref{inversion}),
and (\ref{symmetry}).

We have calculated
the projective factor for the case of genus three surface with
one puncture.
The reason for considering the genus three surface is due to the fact
that for lower genus the projective factor can be absorbed into a
redefinition of generators of the mapping class group. Only starting from
genus three one has simultaneously the lantern~(\ref{lantern}) and
chain~(\ref{chain}) relations for the Dehn twists along non-separating
curves \footnote{the relative normalization of Dehn twists along
non-separating curves is fixed by the braid type
relations~(\ref{braid})}, which are not homogeneous in generators.

The result, given by eqns~(\ref{eqproj}), (\ref{eqcharge}),
is the exponentiated Liouville central charge.
This is in agreement with the connection of the $SL(2,\REALS)$
Chern--Simons and Liouville theories on the classical level, as well as
with the expected interpretation of the Hilbert space of states
in quantum Teichm\"uller theory as the space of Virasoro conformal blocks.
Note that the right answer has been obtained only after
elimination of the non-physical (Gaussian) degrees of freedom
associated with the homologies of the surface.

\section{Acknowledgements}

It is a pleasure to thank L.D. Faddeev for encouragement and
helpful suggestions at various stages of the work. I would also
like to thank
V. Fock, N. Reshetikhin, S. Shatashvili, V. Tarasov,
O. Tirkkonen, S.L. Woronowicz, P. Zograf for discussions.
The work is supported in part by grant RFFI-96-01-00851.

\end{document}